\documentclass[aps,graphicx,twocolumn]{revtex4}
\usepackage{float}
\usepackage{graphicx}
\usepackage{amssymb}
\usepackage{amsmath}
\usepackage{multirow}
\usepackage{booktabs}
\usepackage{epstopdf}
\usepackage{subfigure}
\usepackage[colorlinks,linkcolor=blue,citecolor=blue,urlcolor=blue]{hyperref}
\begin{document}
\title{Efficient hyperentanglement-based quantum secret sharing protocol}
\author{Cheng Zhang$^{1}$, Fu-Le Hao$^{1}$,  Shi-Pu Gu$^{1}$, Xing-Fu Wang$^{2}$, Lan Zhou$^{2}$\footnote{Email address: zhoul@njupt.edu.cn}, Yu-Bo Sheng$^{1}$\footnote{Email address: shengyb@njupt.edu.cn} }
\address{
$^1$College of Electronic and Optical Engineering, \& College of Flexible Electronics (Future Technology), Nanjing University of Posts and Telecommunications, Nanjing, 210023, China\\
$^2$College of Science, Nanjing University of Posts and Telecommunications, Nanjing,\\
}

\date{\today}
\begin{abstract}
Quantum secret sharing (QSS) is a typical multipartite cryptographic primitive, which is an important part of quantum communication network. Existing QSS protocols generally require basis selection and matching, which would increase the quantum resource consumption and classical communication round, and also face weak random security vulnerabilities. We propose an efficient hyperentanglement-based QSS protocol without the basis selection, in which the dealer and partners share a polarization-momentum hyperentangled  Greenberger-Horne-Zeilinger (GHZ) state and encode keys in the polarization degree of freedom (DOF). The dealer decodes the transmitted keys relying on the nonlocal hyperentanglement-assisted polarization GHZ state analysis. Our QSS protocol is unconditionally secure in theory. We develop simulation method to estimate its performance in practical environment. Compared with the QSS based on the GHZ state, our protocol has several advantages.
First, it does not require basis selection and can completely distinguish eight polarization GHZ states, which can improve the utilization rate of entanglement resources to 100$\%$ and increase the key generation rate by an order of magnitude. Second, it only requires one round of classical communication, and thus can reduce key generation time by 69.4$\%$. Third, it can eliminate the weak random security vulnerabilities associated with the basis selection. Finally, our protocol only uses linear optical elements, which makes it practically feasible. Our QSS protocol has potential application in future  quantum communication network.
\end{abstract}
\maketitle

\section{Introduction}
 Quantum communication realizes the information transmission based on the principles of quantum mechanics.
 Quantum communication has some important branches, such as  quantum key distribution (QKD) \cite{QKD1,QKD2,QKD3} and  quantum secret sharing (QSS) \cite{QSS1,QSS2}, both of which have developed rapidly. QKD enables two remote participants to share random keys \cite{QKD1,QKD2}.  QSS is a typical multipartite cryptographic primitive and an important part of quantum communication networks. It enables a dealer to distribute subkeys to multiple partners. All the partners can recover the transmitted key by cooperation. Conversely, QSS enables multiple partners to collaboratively transmit keys to the dealer \cite{QSS1,QSS2}.
In addition to quantum communication field, QSS has important applications in quantum information processing field, such as distributed
quantum computation \cite{computation}.

The first QSS protocol was proposed in 1999 \cite{QSS1}, which is based on the Greenberger-Horne-Zeilinger (GHZ) state. Subsequently, QSS has well developed \cite{QSS2, 2qbQSS1,QSS3,QSSE1,QSS4,QSS5,1qbQSS1,2qbQSS2,QSSE2,2qbQSS3,QSSGraph,1qbQSS2,QSSE3,QSSBE,CVQSS1,QSS6,CVQSS2,QSSE4,QSS7,QSS8,QSSE5,QSSE6,QSSBE2,QSS10,MDI1,MDI2,MDI3,MDI4,MDI5,MDI6,DI1,DI2,DI3,DI4}.
In theory, the two-photon entanglement-based QSS \cite{2qbQSS1}, asymmetric QSS \cite{QSS4}, the single-photon-based QSS \cite{1qbQSS1}, the graph-state-based QSS \cite{QSSGraph}, and the  continuous-variable QSS \cite{CVQSS1,CVQSS2} protocols were successively proposed. In particular, the measurement-device-independent (MDI) and device-independent (DI) QSS protocols were put forward  \cite{MDI1,MDI2,MDI3,MDI4,MDI5,MDI6,DI1,DI2,DI3,DI4}, which can effectively enhance QSS's practical security.
In experiments, Tittel \emph{et al.} realized the experimental demonstration of QSS in 2001 \cite{QSSE1}. In 2013, researchers conducted experiments on phase-encoded QSS in a 50 km optical fiber network \cite{1qbQSS2}. In 2020, an easily scalable QSS was demonstrated \cite{QSSE4}. Recently, finite length phase encoded QSS has been experimentally implemented \cite{QSSE6}.

 The above previous QSS protocols depend on the single-photon measurement, Bell state analysis (BSA), or GHZ state analysis (GSA). However, they all require the basis selection and matching. Only the measurement results in the basis matching cases can be used to generate keys, and those corresponding to the mismatched basis have to be discarded, which leads to a significant waste of precious quantum resources. Meanwhile, implementing the basis matching requires two or more rounds of classical communication, which increases the communication time. As a result, the basis matching requirement largely limits QSS's practical key generation rate. In addition, the basis selection for any participant cannot be completely random under real conditions, which may provide an opportunity for the eavesdropper (Eve) to attack the QSS system  \cite{attack1,attack2}.

Photons have multiple degrees of freedom (DOFs), such as polarization, time-bin, orbital angular momentum (OAM), frequency and spatial mode (momentum). Hyperentanglement is defined as the simultaneous entanglement in two or more DOFs of a quantum system. Hyperentanglement has wide applications for it has great potential to enhance the channel capacity \cite{Capacity1,Capacity2}, improve quantum communication rate \cite{rate}, realize the complete BSA and GSA \cite{CBSA1,CBSA2,GSAsong}, and perform the efficient entanglement purification and concentration \cite{puri1,puri2,puri3,puri4,hyperPS2}. During past few years, various hyperentanglement sources and hyperentanglement applications have developed both in theory and experiment, such as polarization-spatial-mode hyperentanglement \cite{hyper2,hyperPS1,hyperPS3,3zhao,zhao}, polarization-frequency hyperentanglement \cite{hyperPF1,hyperPF2}, polarization-OAM hyperentanglement \cite{hyperPO1,hyperPO2,hyperPO3}, and polarization-time-bin hyperentanglement \cite{hyperPT1,hyperPT2}.

Based on the advantages of hyperentanglement, we propose an efficient hyperentanglement-based three-participant QSS protocol, which does not require basis selection and matching. In our protocol, the dealer Alice and the partners Bob and Charlie share a polarization-momentum hyperentangled GHZ state. All the participants encode in the polarization DOF of their own photons, and then they perform the nonlocal hyperentanglement-assisted (HA) GSA \cite{GSAsong}. Alice can generate the key based on the detector response of Bob and Charlie. Our QSS protocol is unconditionally secure in theory. Compared with previous QSS protocols, it has several advantages. First, it does not require any basis selection and the HA-GSA can completely distinguish eight GHZ states in the polarization DOF, which can improve the utilization rate of quantum resources and increase QSS's key generation rate. In theory, each hyperentangled GHZ state can be used to transmit a 1 bit of key. Second, our protocol only requires \textbf{one} round of classical communication, which effectively reduces the time consumption of the key generation. Compared with the original QSS protocol based on GHZ state \cite{QSS1}, our protocol can reduce the communication time by 69.4\%. Third, our QSS protocol can eliminate the security loophole caused by the basis selection.
 In addition, our protocol completely uses linear optical elements, which is feasible under current experimental conditions. Our QSS protocol can effectively increase QSS's practical key generation rate and enhance QSS practical security, and has application potential in the quantum communication field.

The paper is organized as follows. In Sec. \uppercase\expandafter{\romannumeral2}, we explain our hyperentanglement-based QSS protocol in detail. In Sec. \uppercase\expandafter{\romannumeral3}, we analyze its security and simulate the key generation rate in practical environment. In Sec. \uppercase\expandafter{\romannumeral4}, we give some discussions and draw a conclusion.

\section{Efficient hyperentanglement-based QSS protocol }
Our QSS protocol uses polarization-momentum hyperentanglement. The polarization DOF contains the rectilinear (Z) basis ($Z=\{|H\rangle,|V\rangle\}$) and diagonal (X) basis ($X=\{|+\rangle_{P}=\frac{1}{\sqrt{2}}(|H\rangle+|V\rangle),|-\rangle_{P}=\frac{1}{\sqrt{2}}(|H\rangle-|V\rangle)\}$). Here, $|H\rangle$ and $|V\rangle$ represent the horizontal and vertical polarization states, respectively. The momentum DOF also involves Z basis  ($Z=\{|L\rangle,|R\rangle\}$) and X basis  ($X=\{|+\rangle_{M}=\frac{1}{\sqrt{2}}(|L\rangle+|R\rangle),|-\rangle_{M}=\frac{1}{\sqrt{2}}(|L\rangle-|R\rangle)\}$), where $|L\rangle$ and $|R\rangle$  correspond to distinct momentum states. The illustration of the QSS protocol based on hyperentanglement is shown in Fig. 1, which consists of  the following six steps.
\begin{figure}[ht]
\begin{center}
\includegraphics[width=8.8cm,angle=0]{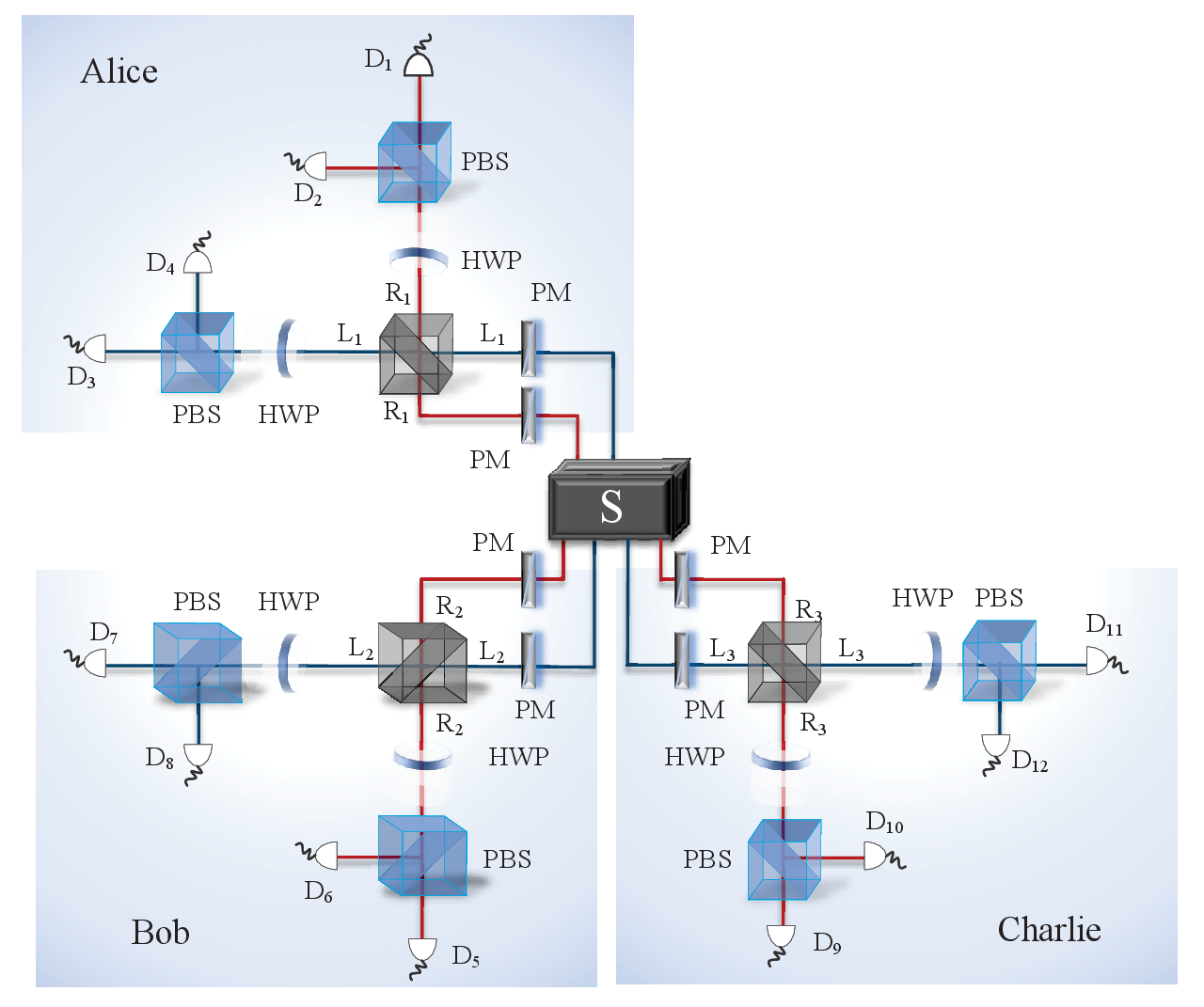}
\caption{Illustration of the hyperentanglement-based QSS protocol. The source S prepares three-photon polarization-momentum hyperentangled GHZ states and sends the three photons to Alice, Bob, and Charlie respectively. PBS (polarization beam splitter) can transmit photons in horizontal polarization $|H\rangle$ and reflect photons in vertical polarization $|V\rangle$. HWP (half wave plate) at $22.5^{\circ}$ can implement Hadamard operation in polarization DOF. PM (polarization modulator) is used to encode information in polarization DOF. $D_{1}-D_{12}$ represent photon detectors, where $D_{2k-1}$ detects the photons in $|H\rangle$ and $D_{2k}$ detects the photons in $|V\rangle$ ($k=1,2,3,4,5,6$).}
\label{fig1}
\end{center}
\end{figure}
\\
\textbf{Step 1}: The hyperentanglement source generates $N$ pairs of identical polarization-momentum hyperentangled GHZ state $|\Phi\rangle=|\psi_{0}^{+} \rangle_{P}\otimes|\psi_{0}^{+} \rangle_{M}$.
The eight GHZ states in the polarization DOF have the forms of
\begin{eqnarray}\label{eq:1}
|\psi_{0}^{\pm} \rangle_{P}=\frac{1}{\sqrt{2}}(|HHH\rangle \pm |VVV\rangle)_{123}, \\
|\psi_{1}^{\pm} \rangle_{P}=\frac{1}{\sqrt{2}}(|VHH\rangle \pm |HVV\rangle)_{123},\nonumber \\
|\psi_{2}^{\pm} \rangle_{P}=\frac{1}{\sqrt{2}}(|HVH\rangle \pm |VHV\rangle)_{123},\nonumber \\
|\psi_{3}^{\pm} \rangle_{P}=\frac{1}{\sqrt{2}}(|HHV\rangle \pm |VVH\rangle)_{123}.\nonumber
\end{eqnarray}
Similarly, the eight GHZ states in the momentum DOF are
\begin{eqnarray}\label{eq:2}
|\psi_{0}^{\pm} \rangle_{M}=\frac{1}{\sqrt{2}}(|LLL\rangle\pm|RRR\rangle)_{123}, \\
|\psi_{1}^{\pm} \rangle_{M}=\frac{1}{\sqrt{2}}(|RLL\rangle\pm|LRR\rangle)_{123},\nonumber \\
|\psi_{2}^{\pm} \rangle_{M}=\frac{1}{\sqrt{2}}(|LRL\rangle\pm|RLR\rangle)_{123},\nonumber \\
|\psi_{3}^{\pm} \rangle_{M}=\frac{1}{\sqrt{2}}(|LLR\rangle\pm|RRL\rangle)_{123}.\nonumber
\end{eqnarray}
Here, the subscript "1", "2", "3" represent the serial number of the photons.

\textbf{Step 2}:  For each hyperentangled GHZ state, the three photons are sent to three participants  through quantum channels. In detail, the photons "1", "2", "3"  are sent to Alice, Bob, and Charlie, respectively.

\textbf{Step 3}: After receiving the transmitted  photons, Alice, Bob, and Charlie randomly encode the photons "1", "2", and "3" in the polarization DOF by  following four unitary operations $\{U_I^P,U_X^P,U_Y^P,U_Z^P\}$, where
\begin{eqnarray}\label{eq:U1}
U_I^P=|H\rangle\langle H|+|V\rangle\langle V|,\quad U_Z^P =|H\rangle\langle H|-|V\rangle\langle V|,  \\
U_X^P =|H\rangle\langle V|+|V\rangle\langle H|, \quad U_Y^P =|H\rangle\langle V|-|V\rangle\langle H|. \nonumber
\end{eqnarray}
Here, $U_I^P$ represents the unitary  operation, $U_X^P$ and $U_Z^P$ represent the bit-flip operation and phase-flip operation, respectively.  $U_Y^P$ stands for the bit and phase flip operation.

It is noticed that these unitary operations are performed only in the polarization DOF, the momentum DOF keeps unchanged. The operations $U_{iB}^P$ of Bob and $U_{iC}^P$ of Charlie together form a one-bit key, that is $k=U_{iB}^P \oplus U_{jC}^P$ $(i,j\in\{I,X,Y,Z\}$).
The specific coding rules in the polarization DOF are as
\begin{eqnarray}\label{eq:3}
&U_I^P=U_Z^P =U_1 \qquad U_X^P=U_Y^P=U_2 \\
&\left\{
\begin{aligned}
U_{iB}\oplus U_{jC}=0,\quad i=j \ (i,j\in \{1,2\})\nonumber\\
U_{iB}\oplus U_{jC}=1,\quad i\neq j \ (i,j\in \{1,2\})
\end{aligned}
\right.
\end{eqnarray}

After the encoding, the GHZ states in polarization DOF shared by the three participants evolve into one of $|\psi_{\mathrm{i}}^{\pm}\rangle_{P}$, $i\in\{0,1,2,3\}$.

\textbf{Step 4}: The photons enter the measurement module. In the measurement module, Alice, Bob and Charlie use the entanglement in the momentum DOF as an auxiliary to perform HA-GSA for each encoded polarization GHZ state \cite{GSAsong}. After the measurement, Bob and Charlie announce their detector responses. In this way, Alice can deduce the encoded GHZ state combined with her own detector response, and thus obtain the key jointly transmitted by Bob and Charlie. Tab. \ref{tab:l} shows the detector responses corresponding to different GHZ states using the momentum state $|\psi_{0}^{+}\rangle_{M}$ as an auxiliary.

\begin{table}
 \centering
    \vspace{-0.3cm}
    \setlength{\abovecaptionskip}{0.3cm}
    \setlength{\belowcaptionskip}{0.1cm}
    \setlength\tabcolsep{10pt}
    \renewcommand\arraystretch{2}

  \caption{The detector response corresponding to different polarized GHZ states using the momentum state $|\psi_{0}^{+}\rangle_{M}$ as an auxiliary \cite{GSAsong}.}
    \begin{tabular}{cl}
         \hline \hline
    \toprule
    \multicolumn{1}{c}{GHZ state} &\multicolumn{1}{c} {Detector response} \\
       \hline
    \multirow{2}[2]{*}{$|\psi_{0}^{+}\rangle_{P}$} & $D_1D_5D_9,D_1D_6D_{10},D_2D_5D_{10},D_2D_6D_9$ \\[-2ex]
          & $D_3D_7D_{11},D_3D_8D_{12},D_4D_7D_{12},D_4D_8D_{11}$ \\
    \midrule
    \multirow{2}[2]{*}{$|\psi_{0}^{-}\rangle_{P}$} & $D_2D_6D_{10},D_2D_5D_9,D_1D_6D_9,D_1D_5D_{10}$ \\[-2ex]
          & $D_4D_8D_{12},D_4D_7D_{11},D_3D_8D_{11},D_3D_7D_{12}$ \\
    \midrule
    \multirow{2}[2]{*}{$|\psi_{1}^{+}\rangle_{P}$} & $D_3D_5D_9,D_3D_6D_{10},D_4D_5D_{10},D_4D_6D_9$ \\[-2ex]
          & $D_1D_7D_{11},D_1D_8D_{12},D_2D_7D_{12},D_2D_8D_{11}$ \\
    \midrule
    \multirow{2}[2]{*}{$|\psi_{1}^{-}\rangle_{P}$} & $D_4D_6D_{10},D_4D_5D_9,D_3D_6D_9,D_3D_5D_{10}$ \\[-2ex]
          & $D_2D_8D_{12},D_2D_7D_{11},D_1D_8D_{11},D_1D_7D_{12}$ \\
    \midrule
    \multirow{2}[2]{*}{$|\psi_{2}^{+}\rangle_{P}$} & $D_1D_7D_9,D_1D_8D_{10},D_2D_7D_{10},D_2D_8D_9$ \\[-2ex]
          & $D_3D_5D_{11},D_3D_6D_{12},D_4D_5D_{12},D_4D_6D_{11}$ \\
    \midrule
    \multirow{2}[2]{*}{$|\psi_{2}^{-}\rangle_{P}$} & $D_2D_8D_{10},D_2D_7D_9,D_1D_8D_9,D_1D_7D_{10}$ \\[-2ex]
          & $D_4D_6D_{12},D_4D_5D_{11},D_3D_6D_{11},D_3D_5D_{12}$ \\
    \midrule
    \multirow{2}[2]{*}{$|\psi_{3}^{+}\rangle_{P}$} & $D_1D_5D_{11},D_1D_6D_{12},D_2D_5D_{12},D_2D_6D_{11}$ \\[-2ex]
          & $D_3D_7D_9,D_3D_8D_{10},D_4D_7D_{10},D_4D_8D_9$ \\
    \midrule
    \multirow{2}[2]{*}{$|\psi_{3}^{-}\rangle_{P}$} & $D_2D_6D_{12},D_2D_5D_{11},D_1D_6D_{11},D_1D_5D_{12}$ \\[-2ex]
          & $D_4D_8D_{10},D_4D_7D_9,D_3D_8D_9,D_3D_7D_{10}$ \\
    \bottomrule
       \hline   \hline
    \end{tabular}
  \label{tab:l}
\end{table}

For example, if Alice performs the $U_{XA}^P$ operation on photon "1", Bob and Charlie perform encoding unitary operations $U_{IB}^{P}$ and $U_{YC}^{P}$ on photons "2" and "3", respectively, the GHZ state in polarization DOF evolves to $|\psi_{2}^{-}\rangle_{P}$, and the encoded GHZ state can be obtained from the responses of the detectors with the assistance of the momentum DOF. By comparing the encoded GHZ state ($|\psi_{1}^{-}\rangle_{P}$) and the initial GHZ state ($|\psi_{0}^{+}\rangle_{P}$) and combining the operation $U_{XA}^P$ imposed by herself, Alice can infer that the operations imposed by Bob and Charlie are ${U_{IB}^{P}}U_{YC}^{P}$ or  ${U_{ZB}^{P}}U_{XC}^{P}$. Both of the operation combinations lead to the $k=0 \oplus 1= 1$.

\textbf{Step 5}: The above processes are repeated until a sufficient number of keys are obtained by Alice. Then, Alice announces a part of keys for the security checking. Three participants announce their operations for the security checking GHZ states. If the measurement results are not consistent with the encoded state, the participants define that an error occurs. After the security checking, if the error rate is higher than the tolerable threshold, the key is considered insecure. All the generated keys should be discarded. If the error rate is lower than the threshold, they proceed to the next step.

\textbf{Step 6}: Alice, Bob and Charlie perform the error correction and private amplification on the transmitted keys to form the final secure keys.

\section{The security analysis and key generation of the hyperentanglement-based QSS protocol}
In this section, we analyze the security of our QSS protocol and prove that our protocol is secure in theory. First, since Alice's encoding operations and measurement results are private, only Alice can deduce the encoded GHZ state and then infer Bob's and Charlie's joint operations. As a result, the external Eve cannot obtain the transmitted keys from the announced detector response.

Second, we analyze the common intercept-resend attack.  In our protocol, the hyperentanglement source generates the identical hyperentangled GHZ states and distribute the photons in each GHZ state to three participants. During the photon transmission process, Eve intercepts the photons sent to Alice. For being detected by Alice, he randomly generates new photons and sends them to Alice. Then, Bob and Charlie perform the encoding operations after receiving photons. After encoding, Eve, Bob and Charlie perform the HA-GSA, and Eve can finally obtain the keys transmitted by Bob and Charlie based on the announced detector responses from Bob and Charlie. However, the above attack can be resisted by the security checking. As the new generated photons by Eve in Alice's location are not entangled with the photons in Bob's and Charlie's locations, they may cause error in the security checking. As long as the number of security checking GHZ states is large enough, it is easy to detect the intercept-resend attack from Eve during the security checking.

Third, we consider the Trojan-Horse attack. During the photon transmission process, Eve can generate some hyperentangled GHZ states and send two photons of each GHZ state to Bob and Charlie. After the encodings, Eve, Bob, and Charlie perform the HA-GSA, and Eve can deduce the transmitted keys from Bob and Charlie. However, as the encoded photons are not sent back to one participants, Eve cannot extract the inserted Trojan photons. The existence of Trojan photons would disturb the order of security checking photon pairs and thus increase the error rate of the security checking. As a result, Eve's Trojan-Horse attack can also be detected by the participants.

Then, we numerically simulate the key generation rate of our QSS protocol.  In our protocol, Bob and Charlie only encode in the polarization DOF and use the momentum entanglement as the auxiliary. According to the Gottesman-Lo-L\"{u}tkenhaus-Preskill (GLLP) theory \cite{GLLP} and secure channel capacity of one-step QSDC \cite{QSDC3}, we can obtain the key generation rate $R_{t}$ of our QSS protocol as
\begin{eqnarray}\label{eq:4}
R_{t} = Q_1^P[1-H(e_1^P)]-Q_t^PfH(E_t^P),
 \end{eqnarray}
where $Q_t^P$ and $Q_1^P$ represent the gain in polarization DOF for all signals and the single pair of polarized GHZ state, respectively, $e_1^P$ and $E_t^P$ represent the error rate corresponding to the single pair of polarized GHZ state and the total bit error rate in the polarization DOF, $f$ is error correction efficiency, and  $H(x)$ represents the binary entropy function with the form of $H(x) = -xlog_{2}(x)-(1-x)log_{2}(1-x)$.

Next, we estimate the above gains and bit error rates. The initial three-photon hyperentangled photon state $|\Phi\rangle= |\psi_{0}^{+} \rangle_{P}\otimes|\psi_{0}^{+}\rangle_{M}$ is generated based on two spontaneous parameter processes. We consider that the hyperentangled GHZ source generates the target hyperentangled photon pair with the probability of $p$. As $p\ll1$, we only consider the vacuum state, one-pair emission, and two-pair emission, and ignore the higher order terms. The practical generated hyperentangled photon state can be written as \cite{zhao}
\begin{eqnarray}\label{eq:6}
\rho=(1-p-p^2)|0\rangle\langle0|+p|\Phi\rangle\langle\Phi|+p^2(|\Phi\rangle\langle\Phi|)^{\otimes2}.
 \end{eqnarray}

After the photon transmission, Alice, Bob, and Charlie share the hyperentangled photon pairs.  Bob and Charlie perform unitary operations to encode keys only in the polarization DOF and Alice, Bob, and Charlie make the nonlocal HA-GSA for decoding. There are totally twelve photon detectors in the nonlocal HA-GSA protocol. As shown in Tab. \ref{tab:l}, each of the eight polarization GHZ states may lead to eight kinds of detector responses with the same probability. We take $|\Phi\rangle=|\psi_{0}^{+} \rangle_{P}\otimes|\psi_{0}^{+} \rangle_{M}$ as an example, which may cause $D_1D_5D_9$, $D_1D_6D_{10}$, $D_2D_5D_{10}$, $D_2D_6D_9$, $D_3D_7D_{11}$, $D_3D_8D_{12}$, $D_4D_7D_{12}$, or $D_4D_8D_{11}$ to respond with the same probability. In our protocol, we consider the practical photon detector which cannot distinguish the number of incident photons and has dark count. We denote $D_ j^k$ as the detection probability of the detector $D_j$ ($j = 1, 2,\cdots, 12$) when $k$ photons are incident. Here, we define the collection efficiency $\alpha=\eta_c\eta_t\eta_d$, which includes the coupling efficiency $\eta_c$ between the nonlinear medium and the fiber, the detection efficiency $\eta_d$ of the detector, and the photon transmission efficiency $\eta_t=10^{(-\beta L/10)}$ ($\beta$ represents optical fiber transmission loss). We assume that the distances $L$ between the hyperentanglement source and Alice, Bob, and Charlie are the same. The dark count rate of the photon detector is given by $p_d$. In this way, we can obtain
\begin{eqnarray}\label{eq:7}
D^k = 1-(1-\frac{\alpha}{4})^k(1-p_d).
 \end{eqnarray}
Here, we define $\alpha' =\frac{\alpha}{4}$, so that we can obtain $D^0=p_d$, $D^1=\alpha'+p_d-\alpha' p_d$, and $D^2=2\alpha' +{\alpha'}^2 p_d+p_d-2\alpha' p_d-{\alpha'}^2$.

According to Eq. (\ref{eq:6}), the hyperentanglement source can generate 0, 1, and 2 hyperentangled photon pairs with different probabilities. In the first scenario, when the photon source generates the vacuum state with the probability of $1-p-p^2$, all the detector clicks are caused by the dark count. As a result, we can calculate $Q_0^P$ as
\begin{eqnarray}\label{eq:8}
Q_0^P& = &(1-p-p^2)\times (D_1^0+D_2^0+D_3^0+D_4^0)\\ \nonumber
&&\times(D_5^0+D_6^0+D_7^0+D_8^0)\\ \nonumber
&&\times(D_9^0+D_{10}^0+D_{11}^0+D_{12}^0) \\ \nonumber
& = &64(1-p-p^2)p_d^3.
 \end{eqnarray}

 In the second scenario, the hyperentanglement source generates one pair of hyperentangled state $|\Phi\rangle$ with the probability of $p$. This hyperentangled photon pair may cause the response of one of the eight detector combinations, say $D_1D_5D_9$, $D_1D_6D_{10}$, $D_2D_5D_{10}$, $D_2D_6D_9$, $D_3D_7D_{11}$, $D_3D_8D_{12}$, $D_4D_7D_{12}$, or $D_4D_8D_{11}$. We suppose that the clicks of $D_1D_5D_9$ are caused by the input photons, and the other detector clicks are caused by the dark count. The number of combinations is $C_{8}^{1}=8$, where $C_{m}^{n}=\frac{m!}{n!(m-n)!}$. In this case, the $Q_1^P$ can be written as
\begin{eqnarray}\label{eq:9}
Q_1^P&=&C_{8}^{1}p\times (D_1^1+D_2^0+D_3^0+D_4^0)\\ \nonumber
&&\times(D_5^1+D_6^0+D_7^0+D_8^0)\\ \nonumber
&&\times(D_9^1+D_{10}^0+D_{11}^0+D_{12}^0) \\ \nonumber
&=&8p(\alpha'+4p_d-\alpha'p_d)^3.
 \end{eqnarray}

In the third scenario, the hyperentanglement source generates two hyperentangled photon pairs with the probability of $p^2$. This situation can be divided into three cases. In the first case, the two photon pairs cause the clicks of one detector combination, i.e. $D_1D_5D_9$ with the probability of $D^2$. The number of combinations is also $C_{8}^{1}=8$. In this case, the gain $Q_{2a}^P$ is
\begin{eqnarray}\label{eq:10}
Q_{2a}^P&=&C_{8}^{1}p^2\times(D_1^2+D_2^0+D_3^0+D_4^0)\\ \nonumber
&&\times(D_5^2+D_6^0+D_7^0+D_8^0)\\ \nonumber
&&\times(D_9^2+D_{10}^0+D_{11}^0+D_{12}^0) \\ \nonumber
&=&8p^2(2\alpha' +{\alpha'}^2 p_d+4p_d-2\alpha' p_d-{\alpha'}^2)^3.
 \end{eqnarray}
In the second case, the two photon pairs  cause the clicks of two detector combinations which are completely uncoincident, i.e. $D_1D_5D_9$ and $D_3D_7D_{11}$. The number of combinations is $C_{4}^{1}C_{4}^{1}=16$. In this case, the gain $Q_{2b}^P$ is
\begin{eqnarray}\label{eq:11}
Q_{2b}^P&=&C_{4}^{1}C_{4}^{1}p^2\times(D_1^1+D_2^0+D_3^1+D_4^0)\\ \nonumber
&&\times(D_5^1+D_6^0+D_7^1+D_8^0)\\ \nonumber
&&\times(D_9^1+D_{10}^0+D_{11}^1+D_{12}^0) \\ \nonumber
&=&16p^2(4\alpha' +2{\alpha'}^2 p_d+3p_d-4\alpha' p_d-2{\alpha'}^2)^3.
 \end{eqnarray}
 In the third case, the two photon pairs cause the click of two detector combinations which are partially coincident, i.e. $D_1D_5D_9$ and $D_1D_6D_{10}$. The number of combinations is $C_{2}^{1}C_{4}^{2}$. In this case, the gain $Q_{2b}^P$ is
\begin{eqnarray}\label{eq:11}
Q_{2c}^P&=&C_{2}^{1}C_{4}^{2}p^2\times(D_1^2+D_2^0+D_3^0+D_4^0)\\ \nonumber
&&\times(D_5^1+D_6^1+D_7^0+D_8^0)\\ \nonumber
&&\times(D_9^1+D_{10}^1+D_{11}^0+D_{12}^0) \\ \nonumber
&=&12p^2(2\alpha' +{\alpha'}^2 p_d+4p_d-2\alpha' p_d-{\alpha'}^2)\\ \nonumber
&& \times(4\alpha' +2{\alpha'}^2 p_d+3p_d-4\alpha' p_d-2{\alpha'}^2)^2.
 \end{eqnarray}
As a result, we can obtain the total gain $Q_2^P$ in the third scenario   as
\begin{eqnarray}\label{eq:12}
Q_{2}^P&=&Q_{2a}^P+Q_{2b}^P+Q_{2c}^P.
 \end{eqnarray}

Then, we consider the case of multiple coincidences, which gives multiple clicks either at Alice's, Bob's, or Charlie's side. Here, we only consider the fourfold click and neglect the events where more than four detectors click simultaneously, for they have much lower probability than the fourfold click. According to above calculations, we can derive the fourfold coincidence rate $Ti$ $(i=0,1,2)$ in above three scenarios as
\begin{eqnarray}\label{eq:13}
T_0&=&(1-p-p^2)(C_{4}^{1})^3C_{9}^{1}p_d^4,   \\ \nonumber
T_1&=&C_{8}^{1}p [C_{9}^{1}(D^1)^3p_d+C_{3}^{2}(D^1)^2 C_{3}^{1}C_{8}^{1}p_d^2\\ \nonumber
&&+C_{3}^{1}D^1(C_{3}^{1})^2C_{7}^{1}p_d^3+(C_{3}^{1})^3C_{6}^{1}p_d^4], \\ \nonumber
T_{2a}&=&C_{8}^{1}p^2[C_{9}^{1}(D^2)^3 p_d+C_{3}^{2}(D^2)^2C_{3}^{1}C_{8}^{1}p_d^2\\ \nonumber
&&+C_{3}^{1}D^2(C_{3}^{1})^2C_{7}^{1}p_d^3+(C_{3}^{1})^3C_{6}^{1}p_d^4], \\ \nonumber
T_{2b}&=&C_{4}^{1}C_{4}^{1}p^2[(C_{2}^{1})^3C_{3}^{1}(D^1)^4+(C_{2}^{1})^3(D^1)^3C_{6}^{1}p_d\\ \nonumber
&&+C_{3}^{2}(C_{2}^{1})^2(D^1)^2C_{2}^{1}C_{5}^{1}p_d^2+C_{3}^{1}C_{2}^{1}(D^1)^2(C_{2}^{1})^2C_{4}^{1}p_d^3\\ \nonumber
&&+(C_{2}^{1})^3C_{3}^{1}p_d^4],\\ \nonumber
T_{2c}&=&C_{2}^{1}C_{4}^{2}p^2[(C_{2}^{1})^2D^2(D^1)^3+(C_{2}^{1})^2D^2(D^1)^2C_{7}^{1}p_d \\ \nonumber
&&+(C_{2}^{1})^2D^2D^1C_{2}^{1}C_{6}^{1}p_d^2+(C_{2}^{1})^2(D^1)^2C_{3}^{1}C_{6}^{1}p_d^2 \\ \nonumber
&&+D^2(C_{2}^{1})^2C_{5}^{1}p_d^3+(C_{2}^{1})^2 D^1C_{3}^{1}C_{2}^{1}C_{5}^{1}p_d^3 \\ \nonumber
&&+C_{3}^{1}(C_{2}^{1})^2C_{4}^{1}p_d^4].
 \end{eqnarray}
In this way, the total fourfold coincidence rate can be calculated as
\begin{eqnarray}\label{eq:14}
T_t=T_{0}+T_{1}+T_{2a}+T_{2b}+T_{2c}.
 \end{eqnarray}

In practical operation, the fourfold click cases should be discarded, so that we can obtain the total gain $Q_t^P$ in Eq. (\ref{eq:4}) as
\begin{eqnarray}\label{eq:15}
Q_t^P=Q_{0}^P+Q_{1}^P+Q_{2a}^P+Q_{2b}^P+Q_{2c}^P-T_t.
 \end{eqnarray}

Next, we calculate the error rate $E_t^P$ in Eq. (\ref{eq:4}). For a specific hyperentangled state, i.e., $|\Phi\rangle=|\psi_{0}^{+} \rangle_{P}\otimes|\psi_{0}^{+}\rangle_{M}$, only the clicks on  $D_1D_5D_9$, $D_1D_6D_{10}$, $D_2D_5D_{10}$, $D_2D_6D_9$, $D_3D_7D_{11}$, $D_3D_8D_{12}$, $D_4D_7D_{12}$, and $D_4D_8D_{11}$ correspond to the correct HA-GSA result, the other items in $(D_1 + D_2 + D_3 + D_4)$ $ (D_5 + D_6 + D_7 + D_8)$ $ (D_9 + D_{10} + D_{11} + D_{12})$ would cause error. In this way, we can calculate the correct gain $Q_{Ci}$ $(i=0,1,2)$ in above three scenarios as
\begin{eqnarray}\label{eq:16}
Q_{C0}&=&(1-p-p^2)C_{8}^{1}p_d^3,   \\
Q_{C1}&=&pC_{8}^{1}[(D^1)^3+3D^1p_d^2+4p_d^3], \nonumber\\
Q_{C2a}&=&p^2 C_{8}^{1}[(D^2)^3+3D^2p_d^2+4p_d^3],\nonumber\\
Q_{C2b}&=&p^2 C_{4}^{1}C_{4}^{1}[2(D^1)^3+6D^1p_d^2],\nonumber\\
Q_{C2c}&=&p^2C_{2}^{1}C_{4}^{2}[2D^2(D^1)^2+2(D^1)^2p_d+4p_d^3].\nonumber
\end{eqnarray}
 In Eq. (\ref{eq:16}), we only consider the influence from the imperfect photon source and the photon detectors on the correct raw key. Actually, during the photon transmission in practical channel and storage in quantum memory, the decoherence in both DOFs are unavoidable. Here, we suppose that the entangled states in two DOFs degrade to the mixed states with the form of
 \begin{eqnarray}\label{eq:17}
\rho_P&=&F_P|\psi_0^+\rangle_P\langle\psi_0^+|+\frac{1-F_P}{7}\left (|\psi_0^-\rangle_P\langle\psi_0^-| \right. \\
&&+|\psi_1^+\rangle_P\langle\psi_1^+|+|\psi_1^-\rangle_P\langle\psi_1^-| +|\psi_2^+\rangle_P\langle\psi_2^+|\nonumber\\
&&\left.+|\psi_2^-\rangle_P\langle\psi_2^-|+|\psi_3^+\rangle_P\langle\psi_3^+|+|\psi_3^-\rangle_P\langle\psi_3^-| \right),\nonumber\\
\nonumber\\
\rho_M&=&F_M|\psi_0^+\rangle_M\langle\psi_0^+|+\frac{1-F_M}{7}\left (|\psi_0^-\rangle_M\langle\psi_0^-| \right.\nonumber\\
&&+|\psi_1^+\rangle_M\langle\psi_1^+|+|\psi_1^-\rangle_M\langle\psi_1^-| +|\psi_2^+\rangle_M\langle\psi_2^+|\nonumber\\
&&\left .+|\psi_2^-\rangle_M\langle\psi_2^-|+|\psi_3^+\rangle_M\langle\psi_3^+|+|\psi_3^-\rangle_M\langle\psi_3^-|\right),\nonumber
 \end{eqnarray}
where $F_P$ and $F_M$ represent the fidelities of the targets entangled states in the polarization and momentum DOFs, respectively.
According to the above formula, we can obtain the total correct gain $Q_{Ct}$ as
 \begin{eqnarray}\label{eq:18}
Q_{Ct}&=&[F_PF_M+\frac{(1-F_P)(1-F_M)}{7}]\\
&&\times(Q_{C0}+Q_{C1}+Q_{C2a}+Q_{C2b}+Q_{C2c}). \nonumber
 \end{eqnarray}
 Therefore, the total error rate $E_t^{P}$ and the error rate $e_1^{P}$ with only a pair of photons can be written as
  \begin{eqnarray}\label{eq:19}
E_t^P=\frac{Q_t^P-Q_{Ct}}{Q_t^P},\\
e_1^P=\frac{Q_1^P-T_1-Q_{C1}}{Q_1^P-T_1}. \nonumber
 \end{eqnarray}
 Taking Eq. (\ref{eq:8}), Eq. (\ref{eq:9}), Eq. (\ref{eq:15}), and Eq. (\ref{eq:19}) in Eq. (\ref{eq:4}), we can finally obtain the value of $R_t$.

\begin{figure}[htbp]
    \begin{center}
   \subfigure[]
 {
        \centering
           \includegraphics[width=8cm,angle=0]{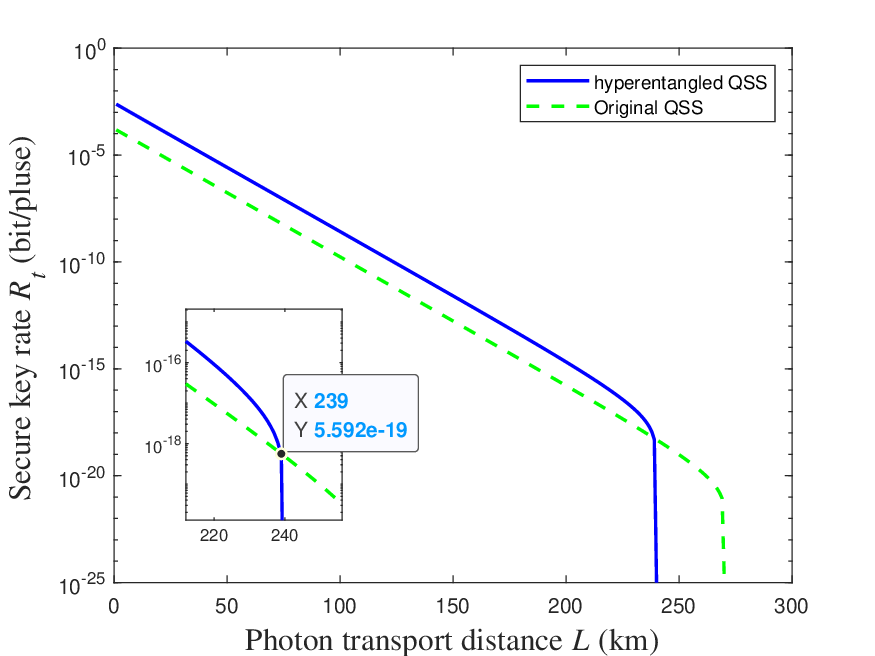}
       \label{fig:R}
 }
    \subfigure[]
 {
    \centering
           \includegraphics[width=8cm,angle=0]{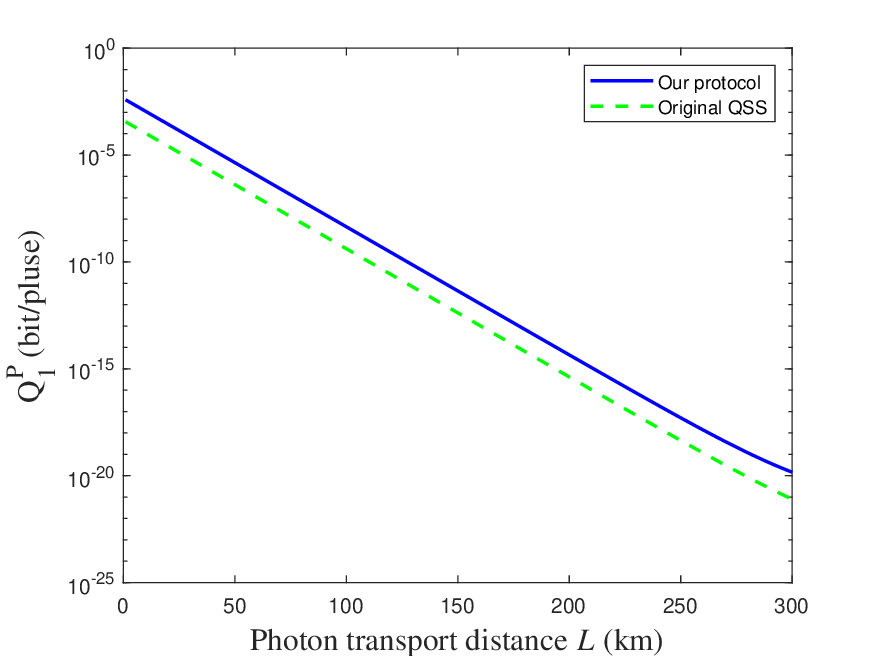}
       \label{fig:error}
  }
   \\
    \subfigure[]
 {
    \centering
           \includegraphics[width=8cm,angle=0]{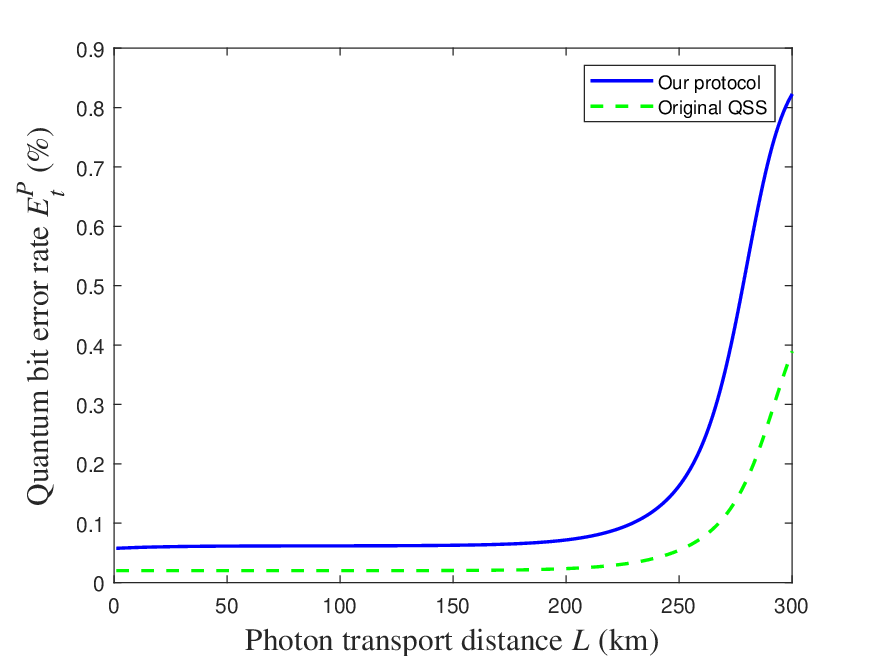}
       \label{fig:Q1}
  }
        \caption{The simulation results of our hyperentanglement-based QSS protocol and the original QSS protocol \cite{QSS1}. (a) The key generation rates of our QSS protocol and the original QSS protocol altered with the photon transmission distance $L$. The small figure in the lower left corner is the enlarged image of the intersection point of the key generation rates, and the data prompt indicates the horizontal and vertical coordinates at the intersection point. (b) The total QBER $E_{t}^{P}$ of our QSS protocol and the original QSS protocol altered with the photon transmission distance $L$. (c) The gain $Q_{1}^{P}$ of our QSS protocol and the original QSS protocol. The simulation parameters are shown in Tab. \ref{tab:parameter}.}
        \label{fig:result}
    \end{center}
\end{figure}

Here, in order to visualize the performance of our protocol, we used some parameters used in Refs. \cite{QSDC3,MDI6} shown in Tab. \ref{tab:parameter} to simulate the protocol, and the simulation results are shown in Fig. \ref{fig:result}.
\begin{table}[ht]
   \centering
  \renewcommand{\arraystretch}{1}
  \caption{Simulation parameters \cite{QSDC3,MDI6}.}
  \setlength{\tabcolsep}{4mm}{
    \begin{tabular}{cccc}
    \hline
    \hline
       $p_d$  & $p$ & $\eta_c$ & $\eta_d $  \\\\[-2ex]
       $10^{-7}$  & $10^{-3}$ & 0.9& 0.93 \\
       \hline
        $\beta $ & $ F_P$ & $F_M$ & $f$\\\\[-2ex]
       0.2 & 0.98 & 0.98 & 1.12\\
         \hline
         \hline
    \end{tabular}
    }
  \label{tab:parameter}%
\end{table}

We compare the key generation rates of the our QSS protocol with the original QSS protocol \cite{QSS1} in Fig. \ref{fig:R}. The results are plotted on a logarithmic scale as a function of the distribution distance $L$ between the source and Alice (Bob and Charlie). It can be found that the maximal photon transmission distance of our QSS protocol is about 239 km, which is lower than that of the original QSS protocol (about 269 km). However, the key generation rate of our QSS protocol is higher than that of the original QSS protocol. In detail, at the photon transmission distance of $L=$ 5 km, 50 km, 100 km, and 200 km, the key generation rate of our QSS protocol is about $1.37\times10^{-3}$, $2.65\times10^{-6}$, $2.633\times10^ {-9}$, and  $2.087\times10^ {-15}$, respectively, while that of the original QSS protocol is about $8.66\times10^ {-5}$, $1.73\times10^ {-7}$, $1.728\times10^ {-10}$, and $1.624\times10^ {-16}$, respectively. In this way, within the scale of maximal photon transmission distance, the key generation rate of the our protocol is about one order of magnitude higher than that of the original QSS protocol.

In order to explain the results of Fig. \ref{fig:R}, we compare the QBER $E_t^{P}$ and the gain $Q_1^{P}$ of the two QSS protocols in Fig. \ref{fig:error} and Fig. \ref{fig:Q1}, respectively. It can be found that the slow growth areas of QBER for both QSS protocols correspond to the uniformly declining area of the key generation rates in Fig. \ref{fig:R}. When $L$ reaches the inflection point (239 km for our protocol and 269 km for the original QSS protocol), the QBER starts to rise rapidly, which lead to the sharply drop of the key generation rate in Fig. \ref{fig:R}. In Fig. \ref{fig:error}, the QBER of our protocol is always higher than that of the original protocol. The reason is that our QSS protocol depends on the entanglement in both DOFs, and the error in any DOF increase $E_t^{P}$. In this way, our QSS protocol is more sensitive for the channel noise. The relatively high QBER of our QSS protocol reduces the maximal photon transmission distance, so that our QSS has shorter maximal photon transmission distance than the original QSS protocol. As shown in Fig. \ref{fig:Q1}, $Q_{1}^{P}$ of the our protocol is always higher than that of the original QSS protocol. The reason is that original QSS protocol requires three participants to perform the single-photon measurement. Three participants' measurement results can be used for generating 1 bit of key only when they select $\{XXX\}$ or $\{XYY\}$ basis combination with the probability of 50\%. In contrast, benefit to the nonlocal HA-GSA with the probability of 100\%, each HA-GSA result can be used to  generate 1 bit of key in theory. In this way, our QSS protocol has higher utilization rate for the entanglement resource, so that it has the higher key generation rate than the original QSS protocol.

\section{Discussion and conclusion}
In the work, we propose \textbf{a} hyperentanglement-based QSS protocol. Besides the key generation rate, it is interesting to compare the communication time required to successfully generate one key bit for our QSS protocol and the original QSS protocol \cite{QSS1}. We assume that the three participants are equidistant from the entanglement source and equidistant from each other, and only consider the photon transmission time and classical communication time. Our QSS protocol requires the source to distribute hyperentangled photons to each participant once with the period of $t_1$, and the detector responses need to be transmitted from Bob (Charlie) to Alice by one round of classical communication with the period of $t_2$. In this way, our QSS protocol takes the time of $t_1 + t_2$ to generate one key bit. In contrast, the original QSS protocol \cite{QSS1} requires one round of GHZ state distribution and two rounds of classical communication (basis selections announcement) to generate one key bit. Meanwhile, as one GHZ state can be used to generate one key bit with the probability of 50\%, generating one key bit needs two pairs of GHZ states in statistic. The total time to generate one key bit for the original QSS protocol is $2t_1$ + $4t_2$. Given that the most time-efficient user distribution is an equilateral triangle with the source at the midpoint, we can obtain $t_2=\sqrt{3}t_1$. In this way, our QSS protocol can save 69.4$\%$ communication time compared to the original QSS protocol. In addition, as our QSS protocol does not require  the basis selection and matching, it can eliminate QSS's weak random security vulnerabilities, and enhance QSS's practical security.

In principle, our protocol can be extended to arbitrary \emph{N}-participant QSS protocol. In theory, one party acts as the dealer and the rest are partners. The source at the midpoint of the N parties generates a N-photon polarization-momentum hyperentangled GHZ state and distributes each photon to a participant. Each of the $N$ participants performs a unitary operation on the received photon. After encoding, $N$ participants perform the HA-GSA, and $N-1$ partners announce their detector responses. The dealer can deduce the key from the published detector response combinations.

Finally, we discuss the experimental realization of our QSS protocol. Our protocol requires the generation and distribution of polarization-momentum hyperentangled GHZ state. Recently, the feasible generation protocol of the hyperentangled GHZ state was proposed \cite{zhao}. By eliminating the constraints of outcome postselection, the generated hyperentangled GHZ state can be directly used in other applications. In the hyperentanglement distribution aspect, the distribution of the polarization-energy-time hyperentanglement via an intra-city free-space link was experimentally realized in 2017 \cite{distribution}. In 2023, the demonstration of the polarization-time-bin hyperentanglement distribution achieves 50 km through noisy fiber channel was reported \cite{distribution1}. Meanwhile, the HA-GSA is also the key element of our QSS protocol. The HA-GSA in Ref. \cite{GSAsong} can completely distinguish  all the eight polarization GHZ states. It only requires the linear optical elements, such as various wave plates and polarization beam splitter, and thus is feasible under current experimental conditions. Based on above progresses, our hyperentanglement-based QSS protocol can be experimentally realized in near future.

In conclusion, we propose an efficient hyperentanglement-based QSS protocol. In the protocol, the hyperentanglement source generates the polarization-momentum hyperentanglement and distributes the photons to three participants Alice, Bob, and Charlie, respectively. The three participants encode the photons only in the polarization DOF and keep the momentum entanglement unchanged. After encoding, they perform the nonlocal complete HA-GSA in polarization DOF assisted by the entanglement in the momentum DOF, and the partners announce their detector responses. The dealer can deduce the encoded polarization GHZ state and decode their transmitted keys. Our protocol is unconditional secure in theory. We develop method to estimate its performance in practical experimental condition. At the photon transmission distance of 5 km, 50 km, 100 km, and 200 km, the key generation rate of our QSS protocol is about $1.37\times10^{-3}$, $2.65\times10^{-6}$, $2.63\times10^ {-9}$, and  $2.09\times10^ {-15}$, respectively. Its maximal photon transmission distance is about 239 km.
 Compared with original QSS protocol based on GHZ state \cite{QSS1}, our QSS protocol has three advantages. First, it does not require the basis selection and matching. The original QSS generates one key bit by using two pairs of GHZ states in statistic, while our QSS protocol can deterministically generate one key bit by using one pair of hyperentangled GHZ state in theory. It leads our protocol to have one order of magnitude higher key generation rate than the original QSS protocol and reduce the entanglement resource by 50\%. Second, our protocol reduces one round of classical communication. In this way, our QSS protocol can save 69.4$\%$ communication time. Third, our QSS protocol can automatically eliminate the weak random security vulnerabilities associated with the basis selection. In addition, our QSS protocol only uses linear optical elements, which is highly feasible under current experimental conditions. Based on above features, our QSS protocol has application potential in the future  quantum communication network.

\section*{Acknowledgement}
This work was supported by the National Natural Science Foundation of
China under Grant Nos. 12175106 and 92365110, and the Postgraduate Research \& Practice Innovation Program of Jiangsu Province under Grant No. KYCX23-0989.

\end{document}